\documentclass[11pt]{article}
\usepackage{graphicx}

\newcommand{\BABARPubYear}    {03}

\newcommand{\BABARConfNumber} {006}
\newcommand{\SLACPubNumber} {9710}
\newcommand{\LANLNumber} {0304020}

\input pubboard/babarsym

\newcommand{\beq}       {\begin{equation}}
\newcommand{\eeq}       {\end{equation}}
\newcommand{\etal}      {{\it et~al.}}

\newcommand{\bsnunu}      {\mbox{$b \to s \nu \overline{\nu}$}}
\newcommand{\Bknunu}      {\mbox{$B^{-}\to K^{-} \nu  \overline{\nu}$}}

\newcommand{\myD}      {\mbox{$\overline{D}^{0}$}}
\newcommand{\myDstar}      {\mbox{$\overline{D}^{*0}$}}

\newcommand{\Dkpi}      {\mbox{$\myD \to K^{+} \pi^-$}}
\newcommand{\DkpipiO}      {\mbox{$\myD \to K^{+} \pi^- \pi^0$}}
\newcommand{\Dkpipipi}      {\mbox{$\myD \to K^{+} \pi^- \pi^+ \pi^-$ }}

\newcommand{\Bsemi}      {\mbox{$B^+ \to \myD X^+$}}

\newcommand{\mybpm}      {\mbox{$B^+B^-$}}

\newcommand{\mybbbar}      {\mbox{$B\overline{B}$}}

\newcommand{\Eextra}       {\mbox{$E_{\rm extra}$}}

\newcommand{\costht}       {\mbox{$|\cos \theta_T|$}}

\newcommand{\ebeam}       {\mbox{$E_{\rm beam}$}}
\newcommand{\ntrks}       {\mbox{$N_{\rm tracks}$}}

\newcommand{\myeffsig}       {\mbox{$\epsilon_{\rm sig}$}}
\newcommand{\myefftag}       {\mbox{$\epsilon_{\rm tag}$}}
\newcommand{\myefftot}       {\mbox{$\epsilon_{\rm tot}$}}
\newcommand{\nsel}       {\mbox{$N_{\rm sel}$}}
\newcommand{\nbg}       {\mbox{$N_{\rm bg}$}}

\newcommand{\eeff}      {\mbox{$e^{+}e^{-} \to f \overline{f}$}}

\newcommand{\pik}      {\mbox{$\pi \to K$}}
\newcommand{\vecp}      {\mbox{$\vec{p}_{\rm B}$}}
\newcommand{\BbkunuSM}       {\mbox{$\mathcal{B}(\Bknunu) = 4 \times 10^{-6}$}}

\newcommand{\LumiOn}       {\mbox{$80.06$~fb$^{-1}$}}
\newcommand{\LumiOff}       {\mbox{$9.6$~fb$^{-1}$}}

\newcommand{\NBBdata}        {\mbox{$(86.9  \pm 1.0) \times 10^{6}$}}
\newcommand{\NBpmdata}       {\mbox{$(43.5  \pm 0.5) \times 10^{6}$}}

\newcommand{\BgTot}          {\mbox{$2.7 \pm 0.7$}}

\newcommand{\BgTotFinal}          {\mbox{$2.7 \pm 0.8$}}

\newcommand{\EffTotFinal}      {\mbox{$(0.046\pm 0.005) \% $}}

\newcommand{\MesSig }       {\mbox{$5.272 \mbox{\rm ~GeV$/c^2$~} < \mes < 5.288 \mbox{\rm ~GeV$/c^2$~}$}}
\newcommand{\MesSide }        {\mbox{$5.225 \mbox{\rm ~GeV$/c^2$~} < \mes < 5.265 \mbox{\rm ~GeV$/c^2$~}$}}

\newcommand{\EextSide }       {\mbox{$ 0.5\mbox{\rm ~GeV~}< \Eextra < 1.5 \mbox{\rm ~GeV}$}}

\newcommand{\NSel }       {\mbox{ 3}}
\newcommand{\BrFinal}       {\mbox{ $\mathcal{B}(\Bknunu) = (0.8 \pm 2.0) \times 10^{-5}$}}

\newcommand{\BrLimit}       {\mbox{ $\mathcal{B}(\Bknunu) < 1.05 \times 10^{-4}$}}
\newcommand{\BrLimitComb}       {\mbox{ $\mathcal{B}(\Bknunu) < 7.0 \times 10^{-5}$}}

\long\def\inst#1{\par\nobreak\kern 4pt\nobreak
    {\it #1}\par\vskip 10pt plus 3pt minus 3pt}

\begin{document}
{\pagestyle{empty}

\begin{flushright}
\babar-CONF-\BABARPubYear/\BABARConfNumber \\
%\babar-PUB-\BABARPubYear/\BABARPubNumber \\
SLAC-PUB-\SLACPubNumber \\
hep-ex/\LANLNumber \\
%March 2003 \\
\end{flushright}

%\begin{flushleft}
%\babar Analysis Document 585 \\
%Version 10 \\
%\today \\
%\end{flushleft}

%\par\vskip 5cm
\par\vskip 4cm

% Title of the paper
\begin{center}
\Large \bf A Search for the Decay $\Bknunu$
\end{center}
\bigskip

\begin{center}
\large The \babar\ Collaboration\\
\mbox{ }\\
\today
\end{center}
\bigskip \bigskip

% Abstract
\begin{center}
\large \bf Abstract
\end{center}
We present a search for the rare flavour-changing neutral-current decay $\Bknunu$ based on 
a sample of $\NBBdata$  $\Upsilon (4S) \to \mybbbar$ events collected in the $\babar$ experiment
at the SLAC $B$-factory.  Signal candidate events are selected by fully
 reconstructing a $\Bsemi$ decay, where $X^+$ represents a combination of up 
to three charged pions or kaons and up to two $\pi^0$ candidates.  The charged tracks and calorimeter clusters not used in the $B^+$ reconstruction are required to be compatible with a $\Bknunu$ decay. We observe a total of three
signal candidate events with an expected background of $\BgTotFinal$, resulting in a preliminary limit of $\BrLimit$ at the $90\%$
confidence level.  This search is combined with the results of a previous and statistically independent 
 preliminary $\babar$ search for $\Bknunu$ to give a limit of  $\BrLimitComb$ at the $90\%$ confidence level.

\vfill
\begin{center}
%Presented at the XVII$^{th}$ Rencontres de la Vall\'ee d'Aoste, \\
%3/9---3/15/2003, La Thuile, Vall\'ee d'Aoste, Italy
Presented at the XXXVIII$^{th}$ Rencontres de Moriond on\\
Electroweak Interactions and Unified Theories, \\
3/15---3/22/2003, Les Arcs, Savoie, France
%Presented at the XXXVIII$^{th}$ Rencontres de Moriond on\\
%QCD and High Energy Hadronic Interactions, \\
%3/22---3/29/2003, Les Arcs, Savoie, France
\end{center}

\vspace{1.0cm}
\begin{center}
{\em Stanford Linear Accelerator Center, Stanford University, 
Stanford, CA 94309} \\ \vspace{0.1cm}\hrule\vspace{0.1cm}
Work supported in part by Department of Energy contract DE-AC03-76SF00515.
\end{center}

\newpage
} % end of pagestyle{empty}

% Input author list file
\begin{center}
\small

The \babar\ Collaboration,
\bigskip

%% author list as of 01-Feb-2003 (555 authors)
%
B.~Aubert,
R.~Barate,
D.~Boutigny,
J.-M.~Gaillard,
A.~Hicheur,
Y.~Karyotakis,
J.~P.~Lees,
P.~Robbe,
V.~Tisserand,
A.~Zghiche
\inst{Laboratoire de Physique des Particules, F-74941 Annecy-le-Vieux, France }
% ...

\end{center}\newpage

% The body of the paper starts here
\section{INTRODUCTION}
\label{sec:Introduction}
\begin{figure}[!h]
\begin{center}
\begin{minipage}{5cm}
\includegraphics[width=5cm]{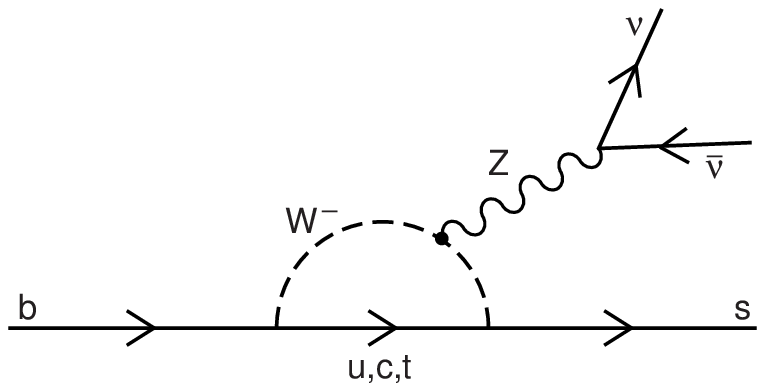}
\end{minipage}
\begin{minipage}{1.0cm}
\mbox{\hspace{1.0cm}}
\end{minipage}
\begin{minipage}{5cm}
\includegraphics[width=5cm]{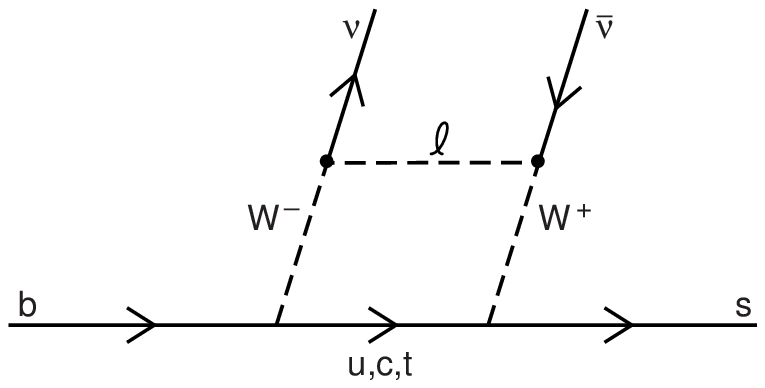}
\end{minipage}
\caption{ Electroweak penguin (left) and box (right) Feynman diagrams for the process $\bsnunu$ predicted by the SM.  In both cases the
amplitudes are expected to be dominated by the heavy $t$ quark contribution.}
\label{fig:feyn}
\end{center}
\end{figure}

The quark level process $\bsnunu$ represents a rare flavour-changing neutral-current (FCNC) decay which
proceeds at the one-loop level in the Standard Model (SM) via ``penguin'' and ``box'' 
diagrams such as those shown in Fig.~\ref{fig:feyn}.
The inclusive $\bsnunu$ process is nearly free from theoretical uncertainties associated
with strong interaction effects, permitting a fairly precise prediction of the SM branching fraction.  The inclusive branching fraction, summed over 
the three neutrino flavours, is estimated to be $(4.1^{+0.8}_{-1.0}) \times 10^{-5}$~\cite{ref:physbook}.  
Since additional heavy particles would also contribute additional loop diagrams, various ``New Physics'' scenarios can
 potentially lead to significant enhancements to the SM branching fraction~\cite{ref:grossman}. 
Unfortunately, an experimental search for the inclusive $\bsnunu$ process is extremely difficult in a $B$-factory environment
 due to the presence of two
unobserved neutrinos which limit the available kinematic constraints that can be exploited in order to suppress
other $B$ decay backgrounds.

Instead, we search for the exclusive $\Bknunu$ decay mode, which proceeds via the $\bsnunu$ process.  
The SM branching fraction for $\Bknunu$ is estimated to be 
$\mathcal{B}(\Bknunu) \simeq 4 \times 10^{-6}$~\cite{ref:buchalla,ref:faessler}.
 The best published limit on the exclusive branching fraction is from CLEO~\cite{ref:cleo} 
with a limit of  $\mathcal{B}(\Bknunu) \leq 2.4 \times 10^{-4} $ at the $90\%$ confidence level.  
 \babar\ has already reported a 
preliminary upper limit $\mathcal{B}(\Bknunu) \leq 9.4 \times 10^{-5} $~\cite{ref:jack} based on $50.7$~fb$^{-1}$ data.
The two $\babar$ analyses use reconstruction methods which produce mutually exclusive data samples,
permitting the two statistically independent results to be combined to obtain an improved limit.

\section{THE \babar\ DETECTOR AND DATASET}
\label{sec:babar}
The data used in this analysis were collected with the \babar\ detector
at the \pep2\ storage ring during the period from 2000 -- 2002, corresponding to a total
integrated luminosity of $\LumiOn$ collected at the $\Upsilon (4S)$ resonance.
This ``onpeak'' sample is estimated to contain $\NBBdata$ $\mybbbar$ pairs.  
This data set is supplemented by a sample of $\LumiOff$ of ``offpeak'' data collected 
approximately $40$~MeV below the $\Upsilon (4S)$ resonance, which is used 
to study continuum background sources due to $\eeff$ where $f = u, d, s, c, e, \mu, \tau$.  

The \babar\ detector is an hermetic detector optimized to provide precision vertexing, charged 
and neutral particle reconstruction and particle identification in an asymmetric $B$-factory
environment.  Tracking
is provided by a five-layer double-sided silicon vertex tracker (SVT), surrounded by a $40$-layer
drift chamber (DCH) filled with a mixture of helium and isobutane.   The SVT and DCH
are situated within a $1.5$~T solenoidal field.
$K$/$\pi$ separation is 
provided by a quartz ring-imaging Cherenkov detector (DIRC) located immediately
outside of the DCH. 
The electromagnetic calorimeter (EMC) is used to measure  energy and position
of photons and electrons.  Muon identification is achieved through segmentation
and instrumentation of the iron of the magnetic flux return (IFR) using resistive plate chambers.  
A more detailed description of the \babar\ detector can be found in~\cite{ref:babar}.

A GEANT4~\cite{ref:geant4} based Monte Carlo (MC) simulation is used to model the signal efficiency and physics backgrounds.  
MC samples equivalent to approximately three times the data luminosity were used to model $B\bar{B}$ events, and samples
equivalent to approximately 1.5 times the data luminosity were used to model continuum events.

\section{ANALYSIS METHOD}
\label{sec:Analysis}

Due to the presence of two unobserved neutrinos, the $\Bknunu$ decay mode~\footnote{Charge
 conjugate modes are implied throughout this paper, however the signal mode will always
be denoted as a $B^-$ decay, while the fully reconstructed tag $B$ will be denoted as a $B^+$ decay to avoid
 confusion.} lacks the kinematic constraints which
are usually exploited in $B$ decay searches in order to reject both continuum and $\mybbbar$ backgrounds.  
Consequently, the approach which is used in this analysis is to first reconstruct the accompanying ``tag'' $B^+$, which 
is produced in association with the signal $B^-$ through the process $\Upsilon(4S)\to \mybpm$,
and then search for evidence of a $\Bknunu$ decay among the tracks and clusters 
not associated with the reconstructed tag $B$. In order to avoid experimenter bias, the signal region in data is not examined (``blinded") until the cuts are finalized.

We reconstruct the tag $B^+$ in a set of decay modes $B^+ \to \myD X^+$ where $X^+$ is a hadronic system composed of up to 
three charged mesons (either $\pi$ or $K$) and up to two $\pi^0$ candidates.  The $\myD$ candidate is 
reconstructed in one of the three decay modes $\Dkpi$, $\DkpipiO$ or $\Dkpipipi$.  Candidate $B^+$ decays are identified 
by combining $\myD$ candidates with sets of charged tracks and $\pi^0$ candidates until the combination yields a $B$ candidate
consistent with the kinematics expected for a true $B$ meson decay.  We use the two kinematic variables $\mes$ and $\Delta E$, defined
by $m_{ES} \equiv \sqrt{{\ebeam}^{2}-{\vecp}^{2}}$ where $\vecp$ is the momentum vector of the
$B$ candidate and $\ebeam$ is the beam energy, and
 $\Delta E \equiv E_{B}-\ebeam$, where $E_{B}$ is the energy of the $B$ candidate.  All these quantities are evaluated in the center of mass (CM) frame. 
If multiple $B$ candidates are identified within the kinematic region $\mes >5.2$~GeV$/c^2$ and $-1.8 {\rm ~GeV~} < \Delta E < 0.6$~GeV, only 
the candidate for which $\Delta E$ is closest to zero is retained.

\begin{figure}[!tb]
\begin{center}
\includegraphics[width=11cm]{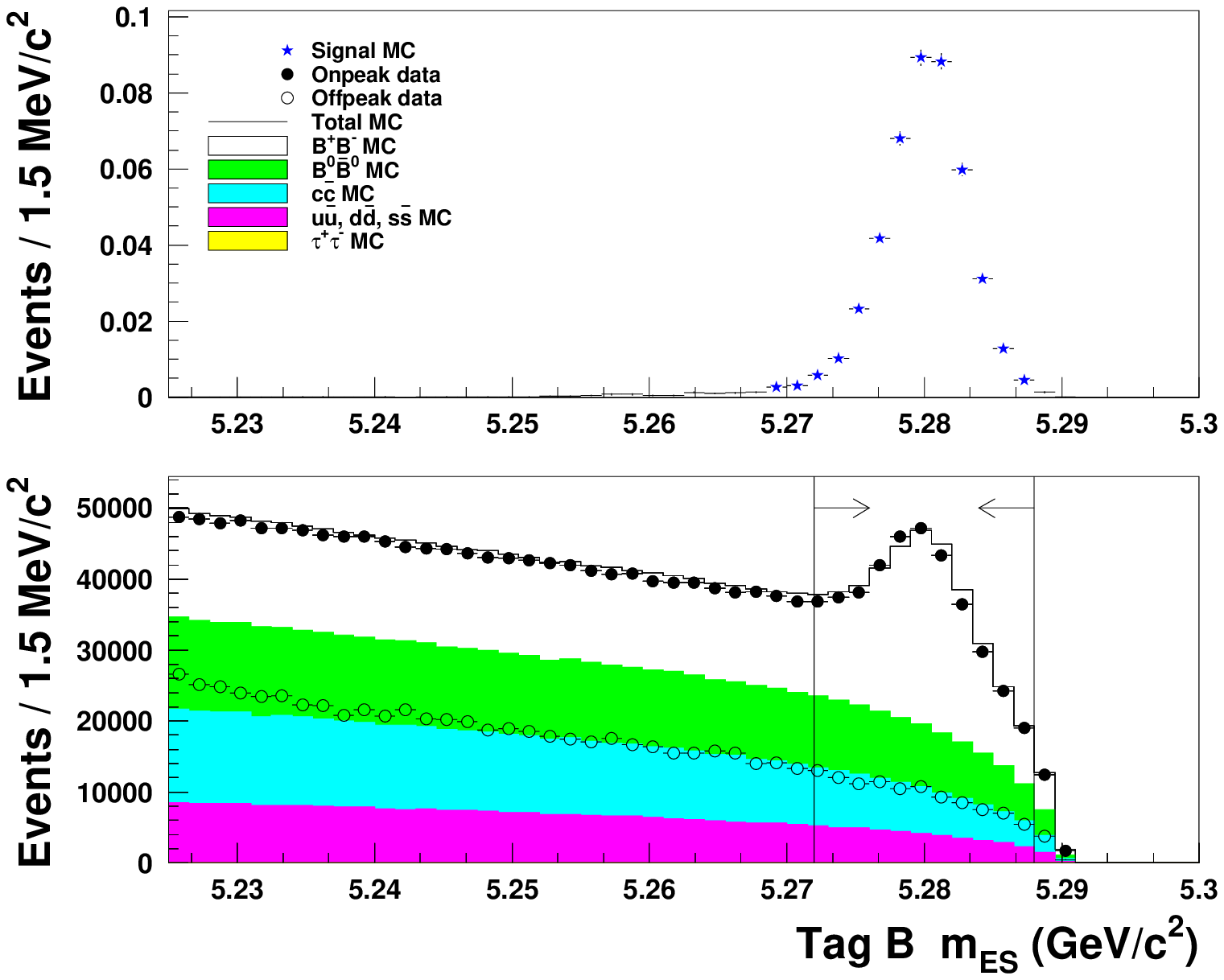}
\caption{The distribution $\mes$ for $B^+ \to \myD X^+$ candidates in the $\Bknunu$ signal MC (top), 
and in the data (bottom). The bottom plot also shows the expected contributions from continuum and $B\bar{B}$ MC. No signal-side selection cuts have been applied.  Signal MC is shown scaled 
to the data luminosity assuming $\BbkunuSM$; inclusive background MC and offpeak data are shown scaled to
 the onpeak data luminosity of $\LumiOn$.}
\label{fig:tag7}
\end{center}
\end{figure}

Combinatorial backgrounds from continuum processes are significantly reduced by 
requiring the thrust, computed using all tracks and clusters in the event, be less than 0.925 and that $\costht$, the
 magnitude of the cosine of  the angle between the thrust axis defined by tracks and clusters used to reconstruct the tag $B$ candidate
 and the thrust axis defined by all other tracks and clusters in the event, be less than 0.8.  
Correctly reconstructed $B$ meson candidates produce a peak in the $\mes$ distribution
above a combinatorial background at the nominal $B$ mass as shown in Fig.~\ref{fig:tag7}.
  The tag $B$ yield is determined directly from data by determining the peaking component of the $m_{ES}$ distribution.
Tag $B$ candidates which are reconstructed in MC events containing a true $\Bknunu$ decay are found to possess very little 
combinatoric background, since there are few additional tracks and clusters in the event which can be randomly combined to produce
combinatoric tag $B$ candidates.    
Events with tag $B$ candidates lying within a signal region defined by $\MesSig$ are retained for use in the search for $\Bknunu$ decays.
Events in the region $\MesSide$ are retained for use in background studies as discussed below.   
A discrepancy of approximately $25\%$ between the observed yield in data and the predicted yield in MC is corrected by scaling the
 peaking component of the MC simulation. 
Good agreement between data and MC is obtained once this correction is applied. The same scale factor, $(75 \pm 7)\%$, is used to correct the tag $B$ yield
 in signal MC as well as MC estimates of peaking backgrounds. The quoted error of $7\%$ is due to the uncertainty in the estimation of the peaking component
 of the data and due to the discrepancy between MC and data in the shape of the non-peaking background component.

\begin{figure}[!tb]
\begin{center}
\includegraphics[width=11cm]{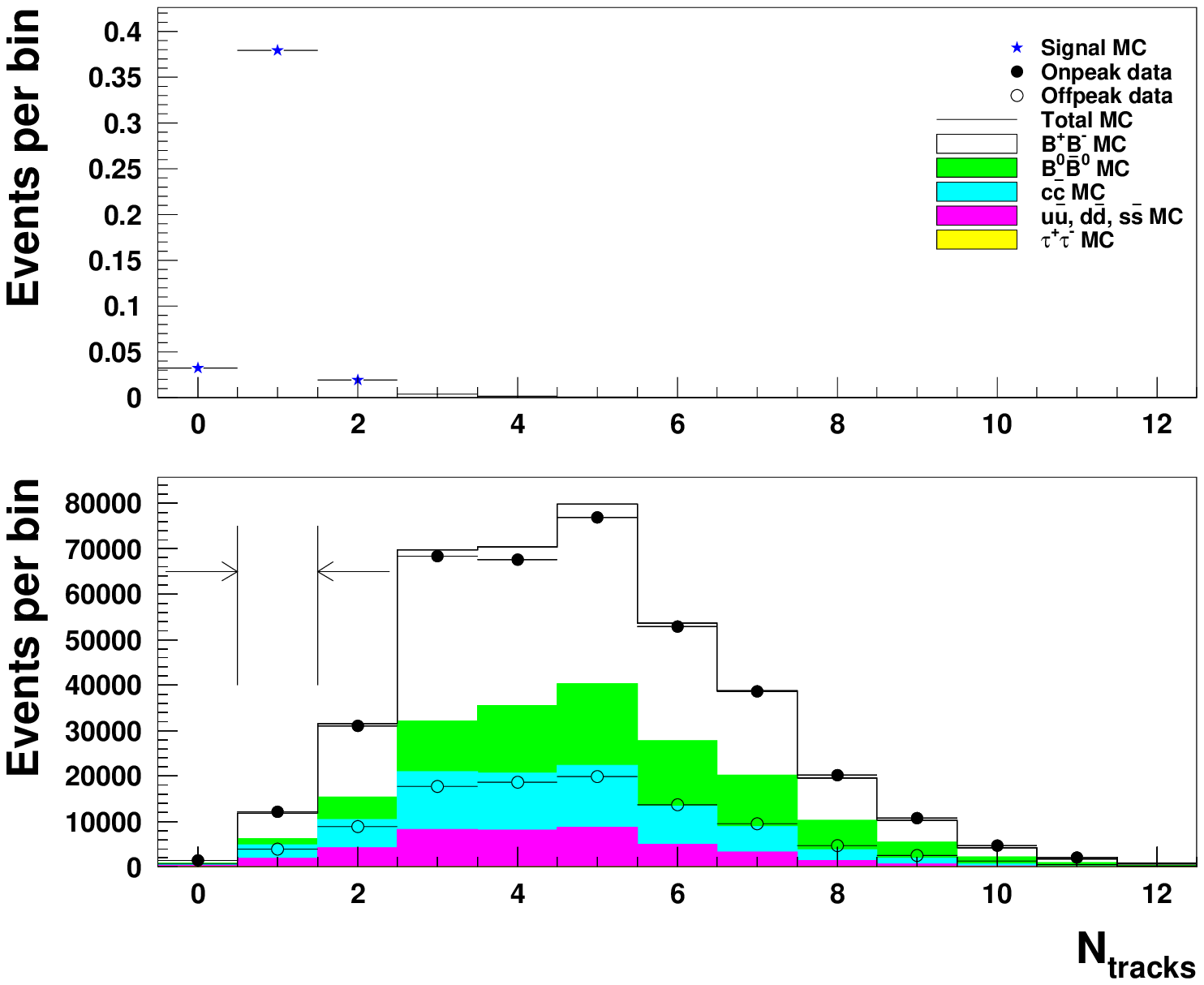}
\caption{ The distribution of the number, $\ntrks$, of signal-side charged tracks is plotted for the 
$\Bknunu$ signal MC (top), and for onpeak data and generic MC (bottom) for events which pass the tag $B$ selection.
No signal-side selection cuts have been applied.
Signal MC is shown scaled to the data luminosity assuming $\BbkunuSM$, and generic MC and offpeak data are shown scaled to
 the onpeak data luminosity of $\LumiOn$. }
\label{fig:sig1}
\end{center}
\end{figure}

\begin{figure}[!tb]
\begin{center}
\includegraphics[width=11cm]{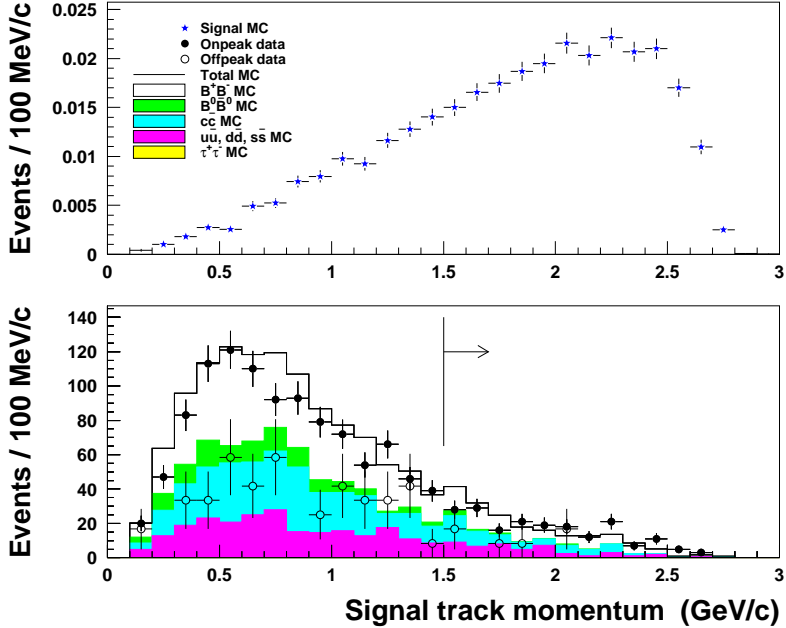}
\caption{ The CM momentum distribution of signal-side kaons candidates is plotted for the 
$\Bknunu$ signal MC (top), and for onpeak data and generic MC (bottom).  Plotted events are required to possess exactly one signal-side track satisfying
kaon identification criteria and having a charge opposite that of the tag $B$.
Signal MC is shown scaled to the data luminosity assuming $\BbkunuSM$, and generic MC and offpeak data are shown scaled to the onpeak data luminosity of $\LumiOn$.}
\label{fig:sig10}
\end{center}
\end{figure}

Once a reconstructed tag $B$ candidate has been identified, $\Bknunu$ signal candidates are selected by considering all tracks and clusters 
in the event which are not used in the tag $B$ reconstruction.  This set of tracks and clusters is referred to in the following as the ``signal-side''
of the event.  Candidate events are required to possess exactly one signal-side reconstructed charged track 
with a charge which is opposite that of the tag $B$.  The signal-side charged track multiplicity is plotted for signal MC and data in Fig.~\ref{fig:sig1}.
The signal candidate track is required to satisfy particle identification criteria for charged kaons based on information from the tracking 
system and the DIRC.  The kaon
candidate is boosted into the CM frame assuming a kaon mass hypothesis, and the CM momentum, $p^*_K$, is required to be greater than $1.5$~GeV$/c$.  The average
particle identification efficiency in the momentum range of interest is $\sim 85 \%$ and the typical $\pik$ misidentification rate is $\sim 2 \%$.  The $p^*_K$
of signal candidate tracks is plotted in Fig.~\ref{fig:sig10}.  
We assume the kaon momentum spectrum is described by the decay model of reference~\cite{ref:buchalla} and correct the signal MC distribution (which is generated 
with a phase-space model) accordingly.  

\begin{figure}[!tb]
\begin{center}
\includegraphics[width=11cm]{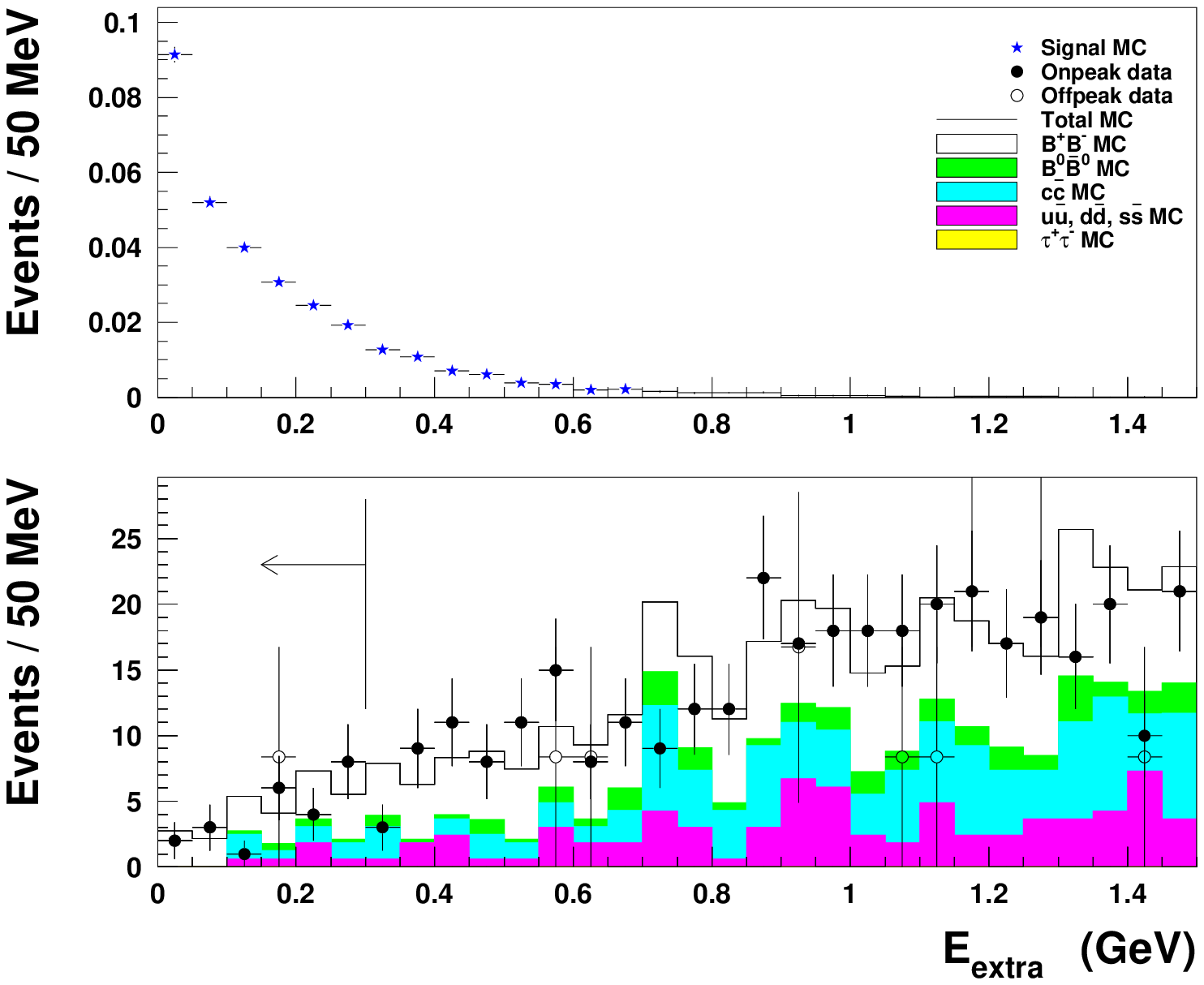}
\caption{ The total extra neutral energy distribution, $\Eextra$ is plotted for the 
$\Bknunu$ signal MC (top), and for onpeak data and generic MC (bottom).  Plotted events are required to possess exactly one signal-side track satisfying
kaon identification criteria and having a charge opposite that of the tag $B$.
 Signal MC is shown scaled to the data luminosity assuming $\BbkunuSM$, and generic MC and offpeak data are shown scaled
 to the onpeak data luminosity of $\LumiOn$.}
\label{fig:sig7}
\end{center}
\end{figure}

Constraints are also imposed on signal-side EMC clusters to reject events with significant neutral energy deposition in the calorimeter.  $\Bknunu$ 
events possess an average of two additional signal-side clusters.  These are generally attributable to hadronic split-offs in the EMC,
usually from pions or kaons associated with the tag $B$ side of the event, or to beam related backgrounds.  Events possessing one or more $\pi^0$ candidates,
composed of combinations of two EMC clusters with lab frame energy greater than $30$~MeV which combine to produce an invariant mass in the range 
$122 \mbox{\rm ~MeV}/c^2 < m_{\gamma \gamma} < 145 \mbox{\rm ~MeV}/c^2$, are rejected.  In addition to this $\pi^0$ veto, a restriction is imposed on 
 the total ``extra'' signal-side neutral energy, $E_{\rm extra}$, that is present in the event.  $E_{\rm extra}$ is computed by summing the CM-frame energies of
all signal-side EMC clusters with lab frame energy greater than $30$~MeV.  Signal candidate events are required to possess $E_{\rm extra}< 300$~MeV.  
The $E_{\rm extra}$ distribution obtained from signal MC is compared to data in Fig.~\ref{fig:sig7}. 

The requirement that signal candidate events possess low signal-side track multiplicity and relatively little additional neutral energy in the EMC
also tends to select $B^+B^-$ events in which the tag $B^+$ has been correctly reconstructed, but the opposing $B^-$ has one or more unreconstructed 
particles which have passed outside the detector acceptance in either the forward or backward directions.  An additional cut is therefore imposed on the 
direction of the missing momentum vector in the CM-frame, $p^*_{\rm miss}$, which is required to satisfy $|\cos \theta_{p^*_{\rm miss}}|< 0.8$.

 Due to the low signal-side multiplicity, there is almost no tag $B$ reconstruction mode-dependence
 of the measured signal-side efficiency.  The overall selection efficiency, $\epsilon_{\rm tot}$,
 can therefore be expressed as the product of the tag $B$ selection efficiency, $\epsilon_{\rm tag}$,
and the signal-side efficiency $\epsilon_{\rm sig}$.  The overall efficiency is estimated to be  $\epsilon_{\rm tot}= \EffTotFinal$, while the
 $\epsilon_{\rm sig}$ is estimated to be $(35 \pm 1)\%$. The uncertainties on the efficiencies are due to both statistics and systematics.   

Monte Carlo modeling of the signal efficiency and background estimates is validated by comparing the yields obtained in a number 
of data control samples.  These samples include $\LumiOff$ of offpeak data, 
an $\mes$ sideband region defined by $\MesSide$, an $\Eextra$ sideband defined by $\EextSide$, and
a ``large $\mes$'' sideband spanning the entire region defined by $\MesSide$ and $\Eextra <1.5$~GeV.
In addition, we retain samples of events which pass all of the nominal signal selection requirements except that they are required to
 have a total of two or three charged tracks associated with the signal side of the event instead of only one.  
All data control samples were found to be in good agreement with the MC predictions, as shown in Table~\ref{tb:bg_estimates}.

\begin{table}[!htb]
\caption{A comparison of data yields and MC predictions in the signal region and in various control samples.
Quoted uncertainties reflect MC statistics only.}
\begin{center}
\begin{tabular}{|c|c|c|c|c|c|c|} \hline
MC type                  & Signal Region  & $\mes$          & large $\mes$    &  $\Eextra$     &  $\ntrks =2$   &  $\ntrks =3$     \\ \hline
$B^+B^-$                 & $1.7\pm 0.6$   & $1.1 \pm 0.5$   & $7.0 \pm 1.4$   &  $3.3 \pm 0.9$ & $17.4 \pm 1.9$ & $54.6 \pm 3.4$   \\
$B^0\overline{B^0}$      &  0             &  0              & $1.4 \pm 0.6$   &  $0.6 \pm 0.4$ & $0.9 \pm 0.5$  &  $3.5 \pm 1.0$   \\
$u\bar{u},d\bar{d},s\bar{s}$              &  0             & $1.8 \pm 1.0$   & $14.0 \pm 2.9$  &  $2.4 \pm 1.2$ & $0.6 \pm 0.6$  &  $1.2 \pm 0.9$   \\
$c\overline{c}$          &  0             & $1.8 \pm 1.0$   & $11.1 \pm 2.6$  &  $2.4 \pm 1.2$ & $1.9 \pm 1.0$  &  $3.1 \pm 1.4$   \\
$\tau^+\tau^-$           &  0             &   0             &  0              &  0             &   0            &   0              \\ \hline
                         &                &                 &                 &                &                &                  \\
Onpeak data              & $\NSel$        &   7             &   31            &  10            &   21           &   55             \\
Total  MC
                 &  $1.7\pm 0.6$  & $4.8\pm 1.7$    & $33.5\pm 4.2$   &  $8.8\pm 2.0$  & $20.7\pm 2.3$  & $62.4\pm 3.9$    \\ 
($\mathcal{L}=\LumiOn$)  &                &                 &                 &                &                &                  \\  \hline
                         &                &                 &                 &                &                &                  \\
Offpeak data             &  0             &   0             &  1              &  0             & 0              &  1               \\ 
Continuum  
 MC             & $0.11 \pm0.05$ &  $0.4\pm 0.2$   & $ 3.0\pm 0.5$   & $0.6\pm 0.2$   & $0.3\pm 0.1$   & $0.5\pm 0.2$     \\
($\mathcal{L}=\LumiOff$) &                &                 &                 &                &                &                  \\ \hline
\end{tabular}
\end{center}
\label{tb:bg_estimates}
\end{table}

Backgrounds consist primarily of $B^+B^-$ events in which the tag $B^+$ has been correctly reconstructed but in which the accompanying $B^-$
decays to a high-momentum kaon and additional particles which are not reconstructed by the tracking detectors or calorimeter.  Typically these
 events contain one or more $K_L^0$ and/or neutrinos, and frequently also one or more additional charged or neutral particles
 which pass outside of the tracking or calorimeter acceptance.  This ``peaking''
background component is evaluated directly from $B^+B^-$ MC and estimated to yield $1.7\pm 0.6$ events in $\LumiOn$ of data.  
This estimate is validated by comparison with the data in the $\Eextra$, $\ntrks =2$ and $\ntrks =3$ sideband, all of which have large contributions 
from peaking $B^+B^-$ backgrounds.  Due to the limited MC statistics, a smaller combinatorial component of the background is estimated by scaling 
the observed yield in the MC $\mes$ 
sideband into the signal region.  This scaling assumes an ``Argus function'' shape for the $\mes$ distribution of the combinatorial component
which is obtained from data using a ``wrong-sign'' tag $B$ sample in which the charge of the $D^0$ daughter kaon is inconsistent with the charge of the tag $B$
candidate, resulting in no significant peaking component in the $\mes$ distribution.  
Scaling the $\mes$ sideband into the signal region yields an additional background of $1.0 \pm 0.4$ events, leading to an estimated total background
of $\BgTot$, where the quoted uncertainties are due to MC statistics.  

\section{SYSTEMATIC STUDIES}
\label{sec:Systematics}

Estimates of systematic uncertainties are summarized in Table~\ref{tb:systematics}.
Systematic uncertainties in the branching ratio determination are dominated by the statistical uncertainty on the background estimate, and
by the uncertainty on the determination of the tag $B$ yield in data using $(75 \pm7)\%$ scaling correction determined from the comparison of $\Bsemi$ yields in data and MC.  Since the tag $B$ yield is determined directly from data, no other 
systematic uncertainties associated with the tag $B$ need to be assigned.  The signal track reconstruction and kaon identification procedure
yield comparatively small uncertainties on the efficiency and background estimates.  Uncertainties also arise from the MC modeling of 
the energy and multiplicity of low energy clusters in the EMC which would potentially produce a bias in the $\Eextra$ distribution.  These uncertainties are
estimated by evaluating the change in the MC efficiency and background estimates when low energy clusters are selectively removed from the 
MC until data and MC multiplicity distributions are in agreement.  An additional systematic uncertainty is assigned to the efficiency estimate as a result
of the re-weighting procedure used to correct the MC kaon momentum spectrum to be consistent with that predicted by reference~\cite{ref:buchalla}.  Note however that
this uncertainty does not account for variations in the momentum spectrum which would result from the use of other theoretical models.   
A small uncertainty also enters the branching ratio limit calculation from the estimation of the number of $B^+B^-$ events present in the 
data sample.  

\begin{table}[!htb]
\caption{
Systematic uncertainties}
\begin{center}
\begin{tabular}{|c|c|} \hline
Source & Relative uncertainty\\ \hline \hline
Signal efficiency:       &   \\
Signal MC statistics $\myeffsig$       &  $1\%$  \\
Signal MC statistics $\myefftag$       &  $5\%$  \\
Tag $B$ yield                          &  $7\%$  \\
Signal track reconstruction efficiency &  $1\%$  \\
$K$ particle ID efficiency                & $2\%$ \\
$\Eextra$                              &  $2\%$  \\ 
Kaon momentum correction               &  $3\%$ \\ \hline
Total  $\sigma \epsilon / \epsilon $   &  $10\%$      \\  \hline \hline
Background estimate:                   &  \\
Generic MC statistics                  &  $27\%$\\
Tag $B$ correction                     & $7\%$  \\
Track efficiency                       & $5\%$ \\ 
$\Eextra$                              & $8\%$   \\ \hline
Total background estimate uncertainty   & $29 \%$ \\  \hline \hline
 $B^+B^-$ yield             &  $1.2 \%$ \\ \hline
\end{tabular}
\end{center}
\label{tb:systematics}
\end{table}

\section{PHYSICS RESULTS}
\label{sec:Physics}

\begin{figure}[!tb]
\begin{center}
\includegraphics[width=13cm]{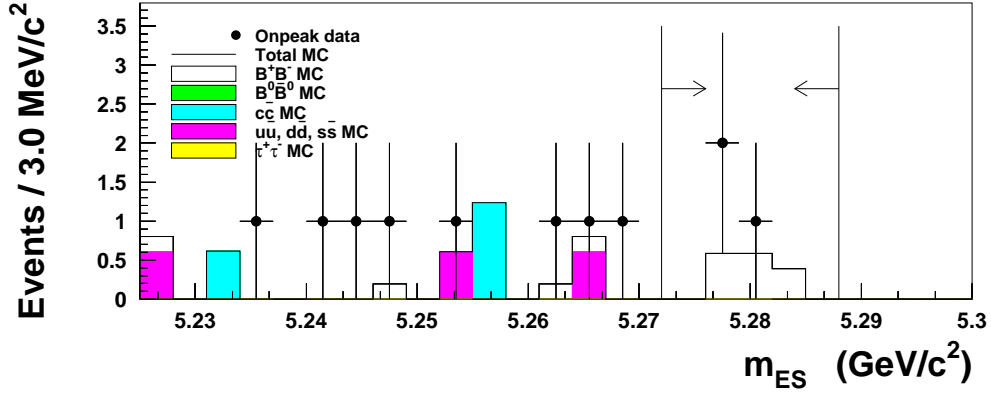}
\caption{The $\mes$ distribution of events passing all other $\Bknunu$ selection cuts, showing the three
selected events in the signal region. }
\label{fig:unblind5}
\end{center}
\end{figure}

\begin{figure}[!tb]
\begin{center}
\includegraphics[width=13cm]{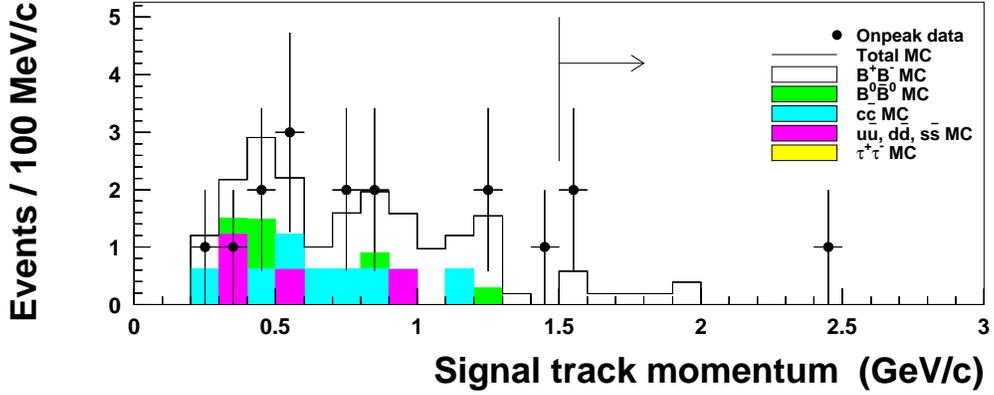}
\caption{The CM momentum distribution of signal kaon candidates which pass all other $\Bknunu$ selection cuts, showing the
three data events above the $1.5$~GeV$/c$ cut. }
\label{fig:unblind6}
\end{center}
\end{figure}

Unblinding the analysis revealed a total of $\nsel = \NSel$ events in the signal region, with an expected background, $\nbg$, of
$\BgTotFinal$ where the additional systematic uncertainties from table~\ref{tb:systematics} have been combined with the 
MC statistical uncertainty. The $\mes$ distribution and signal kaon candidate momentum spectrum are shown in Fig.~\ref{fig:unblind5}
 and Fig.~\ref{fig:unblind6} respectively.  The $\Bknunu$ branching ratio is computed as follows:
\beq
\mathcal{B}(\Bknunu) = \frac{(\nsel - \nbg)}{ 2 \cdot N_{B^+B^-} \cdot \myefftot}   \label{eq:br}
\eeq
where $\myefftot = \EffTotFinal$ is the overall signal selection efficiency, and $N_{B^+B^-}$ is the number of
 $\Upsilon (4S) \to B^+B^-$ events in the data.  $N_{B^+B^-} = \NBpmdata$ is 
obtained by assuming equal branching fractions for $\Upsilon (4S)$ decays into charged and neutral $B$ mesons.  The central value of the branching ratio is 
determined to be $\BrFinal$ where the quoted uncertainty is from systematics only.  The significance of the central value is somewhat less than 
$1 \sigma$ and we therefore quote a $90\%$ confidence level limit.
The branching ratio limit is computed using a frequentist approach based on reference~\cite{ref:cousins}. The confidence level 
for a given branching fraction limit ``guess''
is obtained by generating a large number of experiments in which the systematic uncertainties in the inputs to equation~\ref{eq:br} are modeled by
Gaussian distributions and the signal statistics are modeled by a Poisson distribution.  The limit is set at the value of the branching fraction 
at which $10\%$ of the generated experiments produce a yield which is
less than the observed data yield of three events.   This procedure results in a limit of $\BrLimit$ at the $90\%$ confidence level assuming the central model
of reference~\cite{ref:buchalla}.  This limit is weakly model dependent, since the signal efficiency depends on the kaon momentum spectrum.  Varying the
model within the range specified in~\cite{ref:buchalla} results in a variation in the limit in the range $(1.02 - 1.10) \times 10^{-4}$.  Alternatively, 
the model of reference~\cite{ref:faessler} yields a limit of  $9.5 \times 10^{-5}$.  Sensitivity to new physics will also be model dependent, however it 
should be noted that this analysis has an efficiency which is relatively uniform as a function of kaon momentum above the $1.5$~GeV$/c$ cut.
The efficiency is zero for kaon momenta below $1.5$~GeV$/c$, which may effect the limit interpretation for exotic New Physics modes 
with significantly different kaon momentum spectra.

This analysis can be combined with the result of a preliminary $\babar$ search for $\Bknunu$
reported previously~\cite{ref:jack}, based on a sample of $50.7$~fb$^{-1}$ of data.  
This analysis yielded a limit of 
$\mathcal{B}(\Bknunu) \leq 9.4 \times 10^{-5} $ at the $90\%$ confidence level assuming the model of reference~\cite{ref:buchalla}.
 In this analysis, the tag $B$ was reconstructed in a 
set of semileptonic $B$ decay modes of the form $B^+ \to \myD \ell^+ \nu X^0$ were $X^0$ can be either nothing, or 
one or more photons which are consistent with the decay products of higher mass open charm states such as 
$\myDstar \to \myD \gamma /\pi^0$.  Two events were observed, and were treated as signal for the limit determination 
(i.e. no background subtraction was performed).
Because this previous analysis required an identified high momentum lepton for
the tag $B$ reconstruction, while in the present work the tag $B$ modes are purely hadronic, the two analyses 
are by construction statistically independent and can be readily combined.  Using the same frequentist method as described above and ignoring
correlated systematic uncertainties which are expected to be small, we obtain a combined limit of $\BrLimitComb$ at the $90\%$ confidence level.

\section{SUMMARY}
\label{sec:Summary}
We have performed a search for the rare FCNC decay $\Bknunu$ using a method in which the accompanying tag $B$ meson is reconstructed 
into a set of hadronic final states.  Using a data sample corresponding to an integrated luminosity of $\LumiOn$, we observe a total
of three signal candidate events, consistent with the background expectation of $\BgTotFinal$.  We determine a preliminary limit of the branching fraction 
$\BrLimit$ at the $90\%$ confidence level. Combining our result with the result from a previous and independent $\babar$ search for $\Bknunu$ yields a combined preliminary
limit of $\BrLimitComb$. The result is consistent with the SM 
 expectation of $\mathcal{B}(\Bknunu) \simeq 4 \times 10^{-6}$.

\section{ACKNOWLEDGMENTS}
\label{sec:Acknowledgments}

% Standard acknowledgments paragraph; must always be included.
We are grateful for the 
extraordinary contributions of our \pep2\ colleagues in
achieving the excellent luminosity and machine conditions
that have made this work possible.
The success of this project also relies critically on the 
expertise and dedication of the computing organizations that 
support \babar.
The collaborating institutions wish to thank 
SLAC for its support and the kind hospitality extended to them. 
This work is supported by the
US Department of Energy
and National Science Foundation, the
Natural Sciences and Engineering Research Council (Canada),
Institute of High Energy Physics (China), the
Commissariat \`a l'Energie Atomique and
Institut National de Physique Nucl\'eaire et de Physique des Particules
(France), the
Bundesministerium f\"ur Bildung und Forschung and
Deutsche Forschungsgemeinschaft
(Germany), the
Istituto Nazionale di Fisica Nucleare (Italy),
the Foundation for Fundamental Research on Matter (The Netherlands),
the Research Council of Norway, the
Ministry of Science and Technology of the Russian Federation, and the
Particle Physics and Astronomy Research Council (United Kingdom). 
Individuals have received support from 
the A. P. Sloan Foundation, 
the Research Corporation,
and the Alexander von Humboldt Foundation.

\end{document}